\documentclass[iop,apj]{emulateapj}
\usepackage{amsmath,amssymb,amstext}
\usepackage{subfigure}

\usepackage{environ}
\NewEnviron{myequation}{%
    \begin{equation}
    \scalebox{1.2}{$\BODY$}
    \end{equation}
    }

\usepackage{natbib}
\begin{document}

\title{Clouds in Super-Earth Atmospheres: Chemical Equilibrium Calculations}

\author{Rostom Mbarek}

\affil{Department of Physics, Grinnell College, Grinnell, IA 50112}

\email{mbarekro@grinnell.edu}

\author{Eliza M.-R. Kempton}

\affil{Department of Physics, Grinnell College, Grinnell, IA 50112}

\email{kemptone@grinnell.edu}

\begin{abstract}

Recent studies have unequivocally proven the existence of clouds in super-Earth atmospheres \citep{kreidberg}.  
Here we provide a theoretical context for the formation of super-Earth clouds by determining which condensates are likely to form under the assumption of chemical equilibrium.  
We study super-Earth atmospheres of diverse bulk composition, which are assumed to form by outgassing from a solid core of chondritic material, following \citet{schaef}.  The super-Earth atmospheres that we study arise from planetary cores made up of individual types of chondritic meteorites.  They range from highly reducing to oxidizing and have carbon to oxygen (C:O) ratios that are both subsolar and super-solar, thereby spanning a range of atmospheric composition that is appropriate for low-mass exoplanets.  Given the atomic makeup of these atmospheres, we minimize the global Gibbs free energy of formation for over 550 gases and condensates to obtain the molecular composition of the atmospheres over a temperature range of 350-3,000 K. 
Clouds should form along the temperature-pressure boundaries where the condensed species appear in our calculation.   
We find that the composition of condensate clouds depends strongly on both the H:O and C:O ratios.  For the super-Earth archetype GJ 1214b, KCl and ZnS are the primary cloud-forming condensates at solar composition, in agreement with previous work \citep{gj1214b, kemp2012}.  However, for oxidizing atmospheres, K$_{2}$SO$_{4}$ and ZnO condensates are favored instead, and for carbon-rich atmospheres with super-solar C:O ratios, graphite clouds appear.  For even hotter planets, clouds form from a wide variety of rock-forming and metallic species. 

\end{abstract}

\section{Introduction \label{intro}}

Super-Earth atmospheric observations have received considerable attention as the study of low-mass planets brings us one step closer to eventually characterizing truly Earth-like planets.  
Unlike jovian exoplanets, super-Earths are expected to have diverse atmospheric composition that should strongly depend on the formation history and subsequent evolution of each individual planet.  
Determining the compositions of super-Earth atmospheres has therefore become a priority of exoplanet observers and theorists alike.  
However, the initial observations of super-Earth atmospheres have revealed significant challenges to achieving this goal. Along with the obvious issue -- that the small size of super-Earths makes them especially difficult to characterize -- a second challenge has risen to the forefront:  clouds. 

Clouds do exist in the atmospheres of low-mass exoplanets. Extremely precise transmission spectrum measurements with the WFC3 instrument aboard the Hubble Space Telescope reveal an optically thick high-altitude layer of clouds in the super-Earth GJ~1214b \citep{kreidberg}.  Recently, a second super-Earth, HD~97658b, was observed to have a flat transmission spectrum, consistent with a potential cloud deck \citep{knutson2014}.  Compelling evidence has also been presented that strongly supports the presence of clouds or hazes in a number of Neptune-size exoplanets \citep{knutson, ehr14}. In fact, the only known low-mass exoplanets that have strong evidence for an \textit{absence} of obscuring clouds are the hot Neptunes HAT-P-11b \citep{fra14} and HAT-P-26b \citep{ste16}.
In all of the cases where clouds are believed to be present, the planets' transmission spectra were found to be featureless, implying a gray opacity source high in the planetary atmosphere.  

In many ways, the existence of clouds in super-Earth atmospheres should not be surprising.  Clouds or hazes exist in every solar system body with a substantial atmosphere \citep[e.g.][]{Kra81, Smi82, Ham94, Gri98, Baines2002}. As for gas giant exoplanets, \citet{Iye16} have recently determined that aerosols or clouds block a significant part of the atmospheric column and concluded that cloud layers may be obscuring H$_2$O absorption features in many exoplanet atmospheres. However, it is clear that the study of clouds in exoplanet atmospheres is still in its infancy. While observational studies of exoplanet spectra can reveal the existence of clouds or hazes \citep{pon13, dem13, kreidberg, knutson, ehr14}, a lack of uniquely identifiable spectral features makes it extremely challenging to determine the composition of these clouds from observations alone.  Theoretical studies can fill in the gaps and provide a context for which cloud species might be expected for a given exoplanet atmosphere.  

There are two primary approaches to predict the composition of exoplanet atmospheres -- equilibrium chemistry calculations that determine the composition of an ensemble of gas and condensate species at their lowest global Gibbs free energy state \citep[e.g.][]{1999, Lod02, schaef} and chemical kinetics calculations that follow numerous chemical and photochemical reactions along with vertical mixing of the atmosphere to determine its steady-state composition \citep[e.g.][]{yun82, kas89, hu12}.  The equilibrium chemistry approach has been the primary one used to determine the composition of condensates in planetary atmospheres for two key reasons.  First because the direct minimization of Gibbs free energy allows the complex and difficult-to-model physical details of the condensation process to remain unspecified.  Secondly, the full set of reaction rates and networks that lead to the formation of condensates are often unknown or incomplete making the chemical kinetics approach intractable for cloud studies.  In this paper, we follow the equilibrium chemistry approach to determine the composition of putative clouds in super-Earth atmospheres. 

Thermochemical equilibrium calculations have been a widely used tool for determining the chemical composition of planetary and sub-stellar objects. 
This technique was initially applied to planets of the solar system. For example, \citet{Feg85} and \citet{Vis05} applied this method to fill in gaps in our understanding of the major components of Saturn's atmosphere since observations do not penetrate deep into the atmosphere where key species have condensed out. 
More recently, chemical equilibrium calculations have been applied to brown dwarfs \citep[e.g.][]{Feg94, Feg96, All01, Lod02, Lod06, Cus08, Mor12}, jovian exoplanets \citep[e.g.][]{Vis06, Vis10, Kop12}, and low-mass exoplanets \citep[e.g.][]{gj1214b, Mig14} to study their composition.
Many of these studies also detail the condensed compounds that form in the atmospheres and therefore the clouds.  

Once a basic understanding of the composition of possible cloud layers has been determined from chemical equilibrium calculations, further detailed cloud modeling can be undertaken.  While 3-D models that treat the microphysics of cloud formation along with its coupling to atmospheric radiative transfer is ideal, such models are computationally intensive and have not been broadly applied to exoplanet atmospheres.  Instead, using 1-D models, \citet{Ack01} calculate the vertical profiles of clouds in brown dwarfs and giant planets using an Eulerian framework that parameterizes the cloud scale height and particle size with a single tunable sedimentation parameter.  \citet{Won15} take an alternate approach to determine vertical cloud profiles using a parameterized updraft length scale building off a widely adopted study of equilibrium cloud condensation by \citet{Lew69} that did not treat the vertical cloud structure. \citet{Cah10}, \citet{gj1214b}, and \citet{Mor15} applied the \citet{Ack01} framework to determine the effects of clouds on exoplanet spectra --- the former coupled to a 1-D radiative transfer model to determine the albedo spectra and colors of extrasolar giant planets, and the latter two to super-Earth atmospheres to study the effects of clouds and hazes on the transmission, thermal emission, and scattered light spectra of GJ 1214b.  \citet{For05} also studied the effect of clouds on transmission spectra by determining the abundance of cloud-forming material that would be required to substantially affect the spectrum while accounting for the transit viewing geometry.  

Other detailed models that treat the microphysics of cloud formation have been presented by \citet{Hel08}, \citet{Hel09}, and  \citet{Hel14}. 
These authors introduce a time-dependent description of the formation of stationary dust cloud layers in brown dwarfs and giant planets.
They analyze the formation of such clouds through non-equilibrium processes that include nucleation, growth and evaporation, gravitational settling, and convection.
This microphysical model simultaneously describes the composition of the cloud particles, the grain size, and the size distributions as a function of pressure while also tracking the amount of dust formed and condensable elements still in the gas phase.
This work results in a self-consistent prescription for the particle size distribution and vertical cloud profiles for conditions appropriate to specific sub-stellar and planetary bodies.  Additional work on haze microphysics in Titan's atmosphere by \citet{Lav13} follows the growth of aerosols, focusing on the role of ion chemistry and photochemistry in assembling these large molecules.

Our goal in this paper is to determine the \emph{composition} of cloud-forming materials in a compositionally diverse range super-Earth atmospheres under the assumption of thermochemical equilibrium.  To accomplish this, we must first assume an underlying composition for each atmosphere that we wish to study.  For low-mass exoplanets, degassing during accretion is the likely mechanism for atmosphere formation \citep[e.g.][]{tanton}, since the weak gravitational pull of these planets will inhibit the direct capture of nebular gases.  
We therefore suppose that primitive chondritic material is the source of degassed super-Earth atmospheres.  
A previous paper by \citet{schaef} determined the composition of degassed atmospheres arising from chondritic meteorites of different varieties under the simplifying assumption that each planet is formed from a single type of meteorite. 
Depending on the type of meteorite --- they studied ordinary, carbonaceous, and enstatite chondrites ---  the authors found that the resulting atmosphere could range substantially in bulk composition from H$_{2}$O-rich to H$_2$ or CO-rich.  
Using these atmospheres as a starting point, and seeding them with a small fraction of heavier elements (necessary because \citet{schaef} only followed H, C, N, O, and S in their published results), we determine the composition of condensate clouds that would appear in the eight types of degassed chondritic atmospheres from their study.   

The degassed atmospheres that we consider in this paper span a range of oxidation states with H:O ratios from solar (highly reducing) to less than unity (oxidizing).  They also cover C:O ratios ranging from subsolar to supersolar.  Our models are therefore representative of the diverse range of expected super-Earth atmospheric chemistries. Starting from the atomic composition of the chondritic atmospheres, we next employ a Gibbs free energy minimization routine to determine the composition of
the condensates that appear in the degassed atmospheres across a wide range of pressures and temperatures.  Our calculations account for the rainout phenomenon in which condensed material settles to the lowest location in the atmosphere where it first appears, thereby depleting the upper atmosphere in the atomic species that make up the condensate.  Our end result is to produce condensation curves for the cloud-forming materials in each of the chondritic atmospheres.  Our results are not specifically tied to any particular model atmosphere --- just to the underlying atmospheric composition --- so they are generally applicable to any super-Earth atmosphere in the temperature-pressure range examined in our study.  Our key finding is that the composition of clouds depends strongly on the oxidation state and C:O ratio of the atmosphere in question, as outlined in the following sections.  The remainder of this paper is laid out as follows.  In Section~2, we provide our model description along with a summary of the range of atmospheric compositions that we consider.  In Section~3, we present the results of our chemical equilibrium calculations, including the condensation curves for cloud-forming species. In Section~4, we discuss our results and present some concluding remarks.

\section{Model Description \label{methods}}

\subsection{Chemical Equilibrium Calculations}

Our calculations minimize the Gibbs free energy of formation for more than 550 molecules in the gas and solid phases to establish their abundances in thermochemical equilibrium.
An earlier version of our code (employed in \citet{Kempton2009}) was used to minimize only the Gibbs free energy of gaseous species.
We updated this code to also include condensed species following the methods described in \citet{1990}.
The same methods were revamped in \citet{1999} to include molecules beyond those included in the JANAF thermochemical tables \citep{Cha86}.    

For each atmosphere, our initial condition is a set of atomic abundances for 24 of the most cosmically abundant atoms (see Section~\ref{sec:atom_abun}). 
Our minimization calculation then follows an iterative procedure to determine the final molecular abundances in chemical equilibrium. For each species, the Gibbs free energy of formation $\Delta G_{f}(T)$ is computed with a functional fit to available Gibbs free energy data, of the form:

\begin{myequation}
\Delta G_{f}(T) = aT^{-1}+b+cT+dT^2+eT^3,
\label{energy of formation}
\end{myequation}
where a, b, c, d, and e are the best-fit coefficients.  
The fit coefficients used in this paper are the same ones used in \citet[][Adam Burrows (private communication)]{1999} with the addition of several zinc condensates (ZnS, ZnO, and Zn$_{2}$TiO$_{4}$) whose fit coefficients were calculated specifically for this work based on their tabulated Gibbs free energies of formation from \citet{Rob95}.  The fits are calculated relative to a non-standard zero-point reference state, which is the atomic gas phase of each of the atoms that make up a particular molecule, as described in \citet{1990}.  For each molecule, the fit to the available Gibbs free energy data is known to be valid over a set temperature range, beyond which the molecule is typically excluded from our calculation.  However, we find that there are certain cases in which extrapolations of the Gibbs free energy fits beyond the specified temperature range are necessary when high abundances of those molecules persist all the way up (or down) to the temperature cutoff.

The total equilibrium Gibbs free energy of the entire system is found by minimizing the following function:

\begin{align*}\label{eq:pareto mle2}
\frac{G(T)}{RT} &= \sum_{k=1}^{m} y_k\biggl(\frac{\Delta G_{f}(T)}{RT}+ln(P)+ln\frac{y_k}{y_{sum}}\biggr) \newline 
\end{align*}
\begin{equation}
+ \frac{1}{RT}\sum_{k=m+1}^{s}y_k\Delta G_{f}(T), 
\end{equation}

 where $m$ is the number of the gas species, $s$ is the total number of the gas and
condensed species, $y_k$ is the mole fraction of species $k$, $R$ is the gas
constant, $T$ is the temperature, $P$ is the total pressure, and $y_{sum}$ is the number of moles in the gas phase. 
Our minimization procedure follows the iterative method of steepest decent as laid out in \citet{1958}.  
At each temperature-pressure grid point in our calculation, we consider the solution to be converged when two subsequent steps in the minimization routine do not lead to significant fractional changes in abundance for any of the molecules.  

We perform our fits over a pressure range of $10^{-6} - 10^{3}$ bar and a temperature range of $350 - 3,000$ K.  The low-temperature cutoff of 350 K was chosen because many of our Gibbs free energy fits do not extend below 300 K, and our database is currently lacking data for many types of ices, which are expected to be abundant at lower temperatures.  Across the temperature-pressure range of our calculations our molecular database of Gibbs free energies includes known geophysically and atmospherically relevant species along with many other species made of stoichiometric combinations of the 24 atoms included in our calculations.  Of note, since our calculations extend previous work that focused on near-solar composition atmospheres to much more oxidizing scenarios, we include the more highly oxidized variants on many of the molecules that appear in previously published work.  For example, iron is predicted to condense as solid Fe in solar composition atmospheres, but our database also includes the oxidized iron-bearing molecules FeO, Fe$_{2}$O$_{3}$, and Fe$_{3}$O$_{4}$, along with a number of other iron-bearing condensates that incorporate additional atoms.

\subsection{Atomic Abundances \label{sec:atom_abun}}

Predicting the composition of atmospheres using chemical equilibrium calculations is contingent on understanding the evolution of the rocky Super-Earth planets we are considering.  In our calculations, this arises as a need to specify the atomic composition of the atmosphere as the initial condition for our Gibbs free energy minimization routine.  We specifically study atmospheric compositions that are relevant to outgassed super-Earth atmospheres.   
Here we rely on previous work done by \citet{schaef} who calculated the composition of outgassed atmospheres for planetary interiors composed of different individual types of meteorites.  
Their work looked specifically at 8 types of ordinary, carbonaceous, and enstatite chondrites: CI, CM, CV, EH, EL, H, L, and LL.  The authors explicitly assumed that the atmosphere would be outgassed during accretion from a hot, fully molten, non-differentiated planet, and that the outcome of the outgassing process was well-described by chemical equilibrium calculations.  They tested this second assumption through a limited number of chemical kinetics calculations and found the equilibrium chemistry assumption to be valid.  By using the results of \citet{schaef}, we are implicitly making the same set of assumptions in our own work.   The authors determined the abundances of gas phase H$_{2}$, H$_{2}$O, CH$_{4}$, CO$_{2}$, CO, N$_{2}$, NH$_{3}$, H$_{2}$S, and SO$_{2}$ for atmospheres produced by each type of meteorite, and found that the composition of the atmospheres varied considerably across their models. 

Starting from the molecular composition of these chondritic atmospheres \citep[][Table 1]{schaef}, we determine their atomic makeup by the following method.  We break up the molecules that comprise each atmosphere stoichiometrically to determine the relative abundances of H, C, N, O, and S.  At this point, the total abundances still do not add to 100\%, since \citet{schaef} included a final category of "other" species, which are molecules other than the 9 previously mentioned.  For the purpose of this work, we assume that the "other" category is made up exclusively of atoms other than H, C, N, O, and S.  To populate the remainder of the atmosphere, we calculate the abundances of \textbf{16} additional refractory elements listed in Table~1 in solar composition ratios \citep[using the solar system abundances of][]{Lodders} that would be required to bring the total abundance to 100\%.  Chemical equilibrium considerations imply that even these refractory elements will outgas into the atmosphere at temperatures above their boiling points.  While these species will not be major atmospheric constituents owing to their low abundances, they can still play an important role in forming condensate clouds as our calculations reveal in the following section.  Finally, we set the abundances of the noble gases He, Ne, and Ar (which are also tracked by our code) to zero for the degassed atmospheres. 

The resulting atomic abundances for each of the chondritic atmospheres are reported in Table~1.  It is important to note that while we do not necessarily expect to find super-Earths with the exact abundance ratios of any of the individual classes of chondrites, our approach allows us to study the condensates that will form in outgassed atmospheres of varying H:C:N:O ratios that represent a plausibly diverse set of low-mass planets.    

\begin{figure}
  \centering
  \begin{subfigure}
    \centering
    \includegraphics[angle=90,scale=.4, trim= 90 90 50 150]{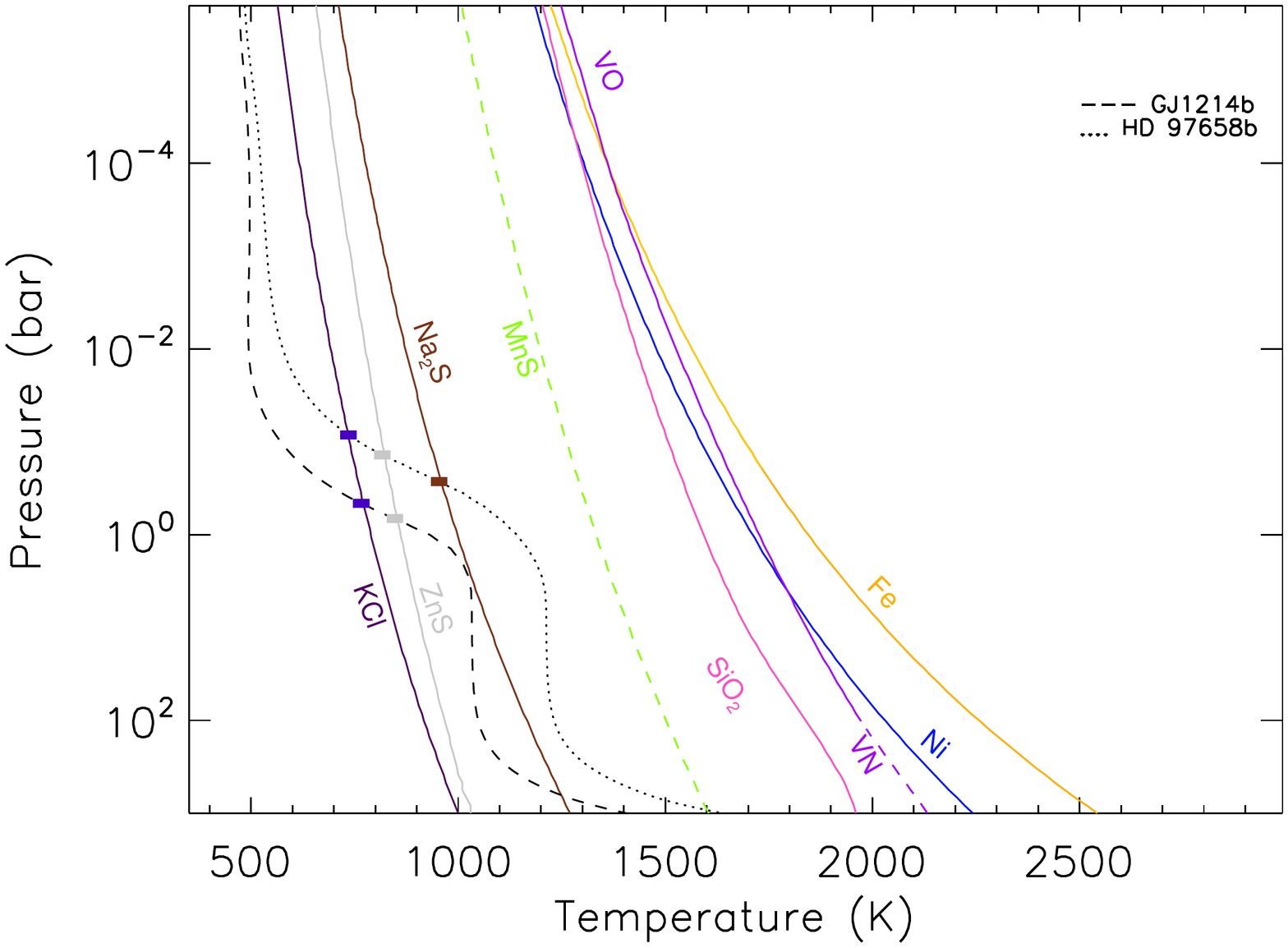}
  \end{subfigure}%
  \quad
  \begin{subfigure}
    \centering
    \includegraphics[angle=90,scale=.4, trim= 90 90 50 150]{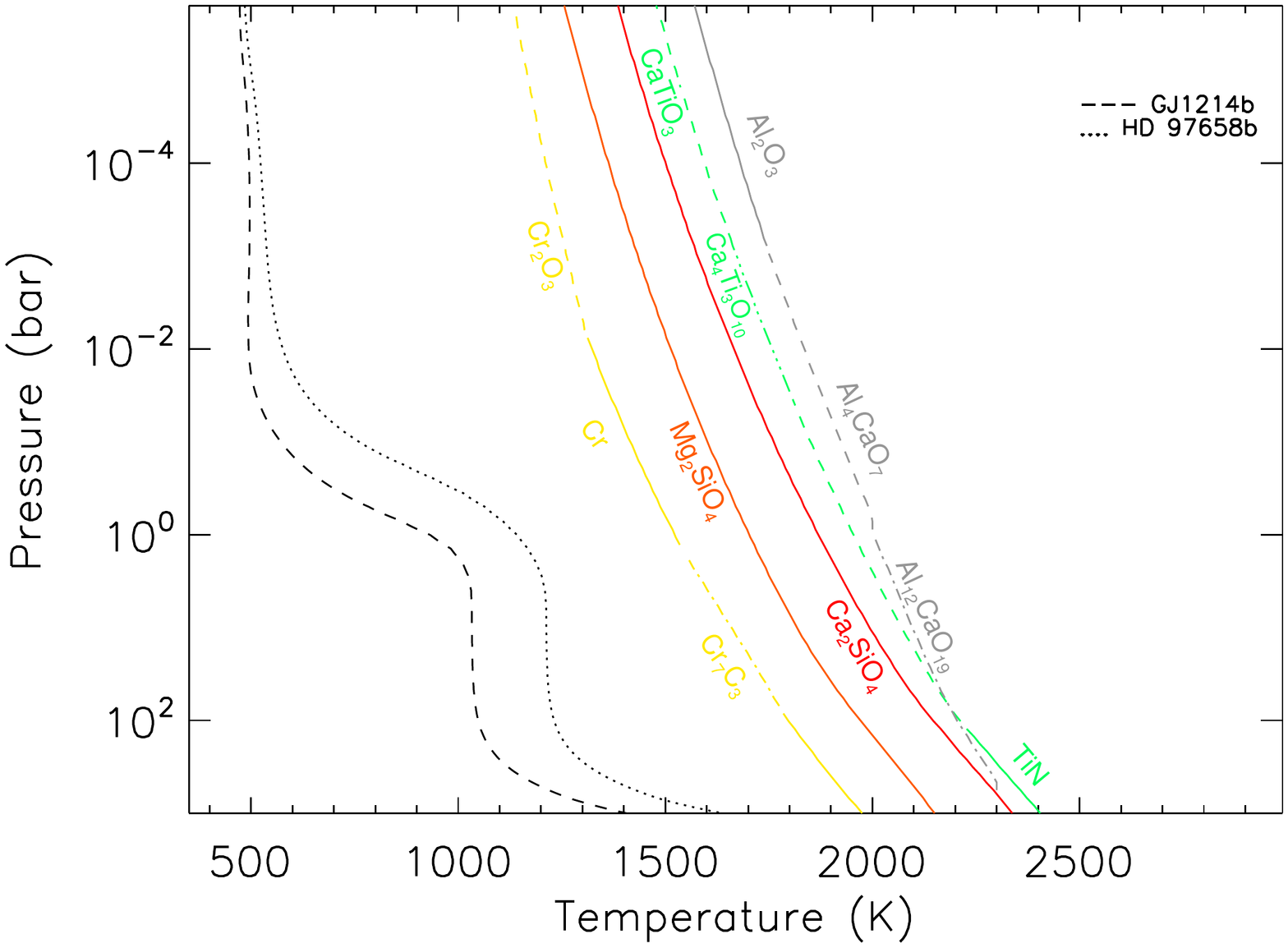}
  \end{subfigure}
  \caption{Condensation curves for an atmosphere of solar composition. The T-P profiles of the super-Earth archetypes GJ 1214b and HD 97658b calculated at solar composition \citep[][and Jonathan Fortney private communication]{mil10} are overlaid for reference.  Intersections of the condensation curves with the T-P profiles indicate the composition and location of putative clouds. The location and composition of the base of the cloud layers are indicated with filled squares.}
  \label{fig:figures1}
\end{figure}

\subsection{Rainout Calculations}

Chemical equilibrium calculations alone are not sufficient to determine the composition of clouds in planetary atmospheres.  Simply minimizing Gibbs free energy as a function of temperature and pressure fails to account for the rainout phenomenon.  As a molecule in the condensed phase appears, it tends to fall within the gravitational field of the planet. This causes the atmosphere to be depleted of that specific molecule (and its atomic constituents) at altitudes above that of the cloud.  The rainout phenomenon therefore alters the composition of the planetary atmosphere above the cloud. Secondary condensates of most atoms will not form because the primary (highest temperature) condensate will draw its constituent atoms down to the location in the atmosphere where the cloud forms.  In the case where the entire atmosphere remains below the condensation temperature of a particular atom, that species will be sequestered in solid or liquid form within the planet's surface and interior.  

To illustrate the effects of the rainout phenomenon, consider the following case. For a straightforward Gibbs free energy minimization calculation of a solar composition ensemble \textit{without} rainout, Fe and FeS both appear as iron condensates \citep[see e.g.][Figure 2]{1999}.  Fe is the primary condensate appearing at high temperature, and FeS is the secondary condensate, which appears around 700 K.  The no-rainout calculation also results in the formation of NaCl condensates within the T-P range of GJ 1214b's atmosphere, rather than KCl clouds that have been reported by authors who do include the rainout effect \citep{Lod06}.  When calculations are performed that account for rainout, the following changes occur.  Fe remains as the primary iron-bearing condensate at solar composition, but FeS does not appear at lower temperatures because iron is sequestered within the Fe cloud at a hotter location lower in the atmosphere \citep[e.g.][]{Vis10}.
As a result, sulfur binds with sodium instead of iron at lower temperatures to form the Na$_2$S condensate. The atmosphere therefore becomes depleted in sodium which in turn makes the formation of the NaCl cloud at even lower temperatures impossible.  We then obtain the KCl cloud (shown in Figure~1) at temperatures that intersect the T-P profile of GJ 1214b, in agreement with previous work.

\begin{figure}
\centering
\includegraphics[angle=90,scale=.4,clip=true, trim= 60 50 50 140]{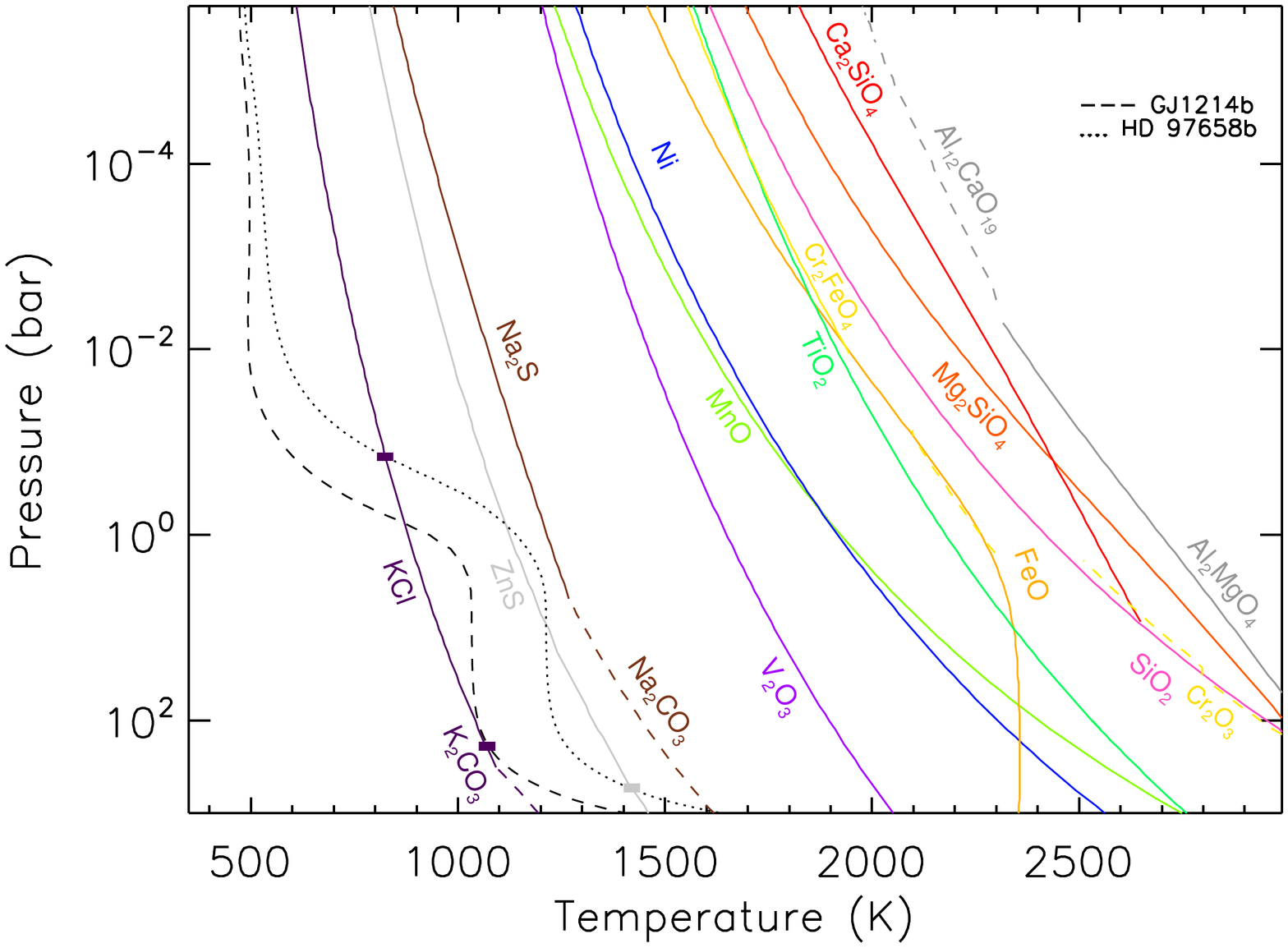}
\caption{Condensation curves and cloud locations for an atmosphere formed from CM chondritic material.}
\label{fig:fig2}
\end{figure}

\begin{figure}
\includegraphics[angle=90,scale=.4,clip=true, trim=  60 50 50 140]{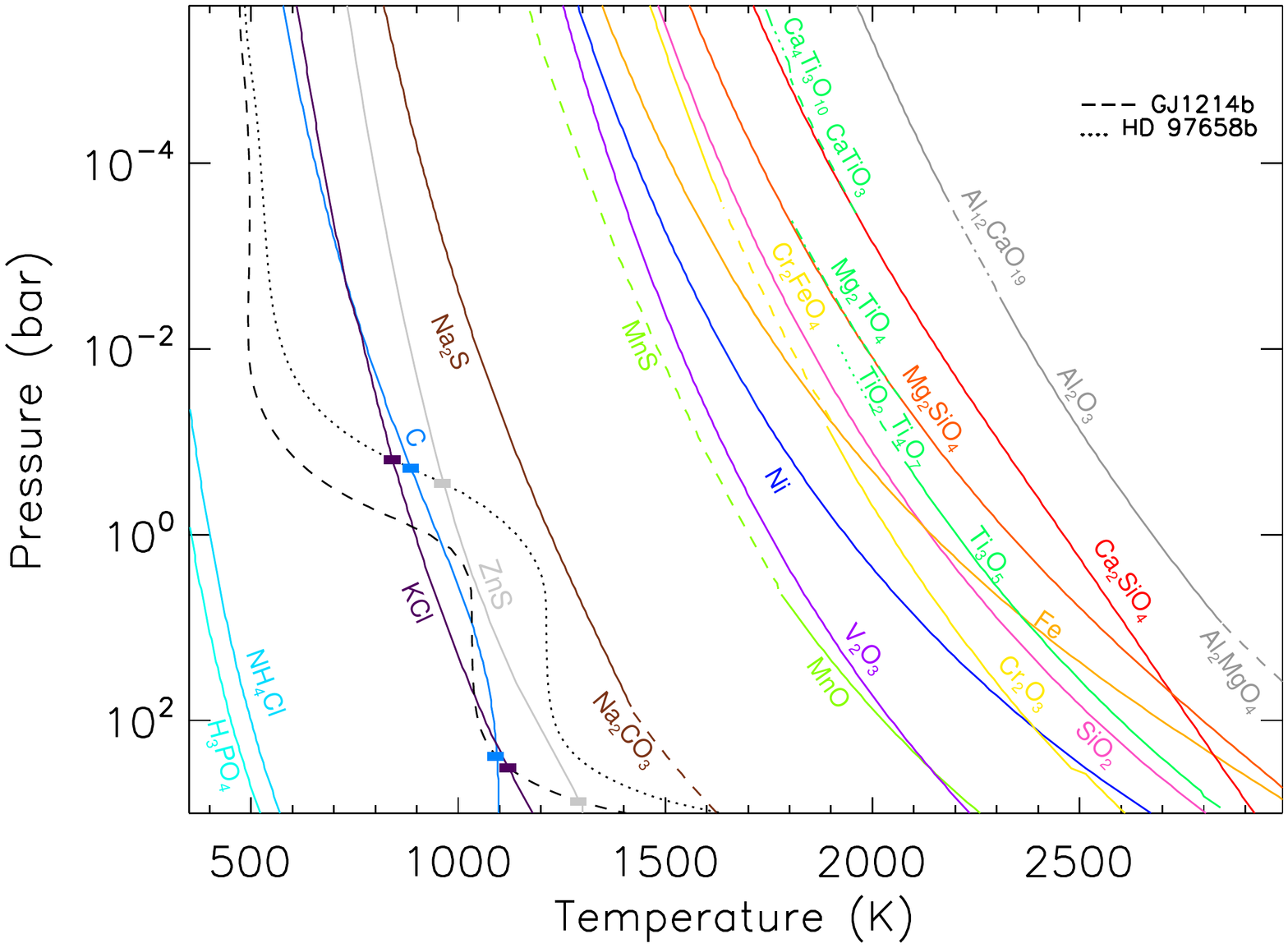}
\caption{Same as Figure~\ref{fig:fig2} but for H chondritic material.}
\label{fig:fig3}
\end{figure}

We account for rainout within the framework of our Gibbs free energy minimization code by removing atoms from our calculation as they are depleted by condensation processes.  Starting at the highest temperature point in our calculation (3,000 K in this case), the composition of the atmosphere is specified by the atomic abundances in Table~1.  
Our code then calculates the molecular abundances at sequentially lower temperatures until the first condensate appears.  At this point the atoms that make up the condensate are subtracted from the initial atomic abundances in their stoichiometric ratios until one of the atoms is fully depleted.  
The updated atomic abundances are then used as the code proceeds to calculate abundances at subsequent lower temperatures.  
As the temperature decreases, the atomic abundances are updated to account for the depletion effect each time a condensed species appears in our calculation. 
The following example will further illustrate this concept. 
Molecule X$_a$Y$_b$Z$_c$ appears as a cloud in our calculation such that the abundances of the atoms X, Y, and Z are respectively A$_X$, A$_Y$, and A$_Z$. Finding the minimum of aA$_X$, bA$_Y$, and cA$_Z$ allows us to determine which atomic species will become fully depleted by the formation of this cloud.  If, for example, X becomes fully depleted, we then update the atomic abundances of X, Y, and Z to zero, A$_Y - (b/a)$A$_X$, and A$_Z - (c/a)$A$_X$, respectively.  The number of atomic species we consider therefore decreases as our calculation progresses to lower temperatures and atoms are removed from the atmosphere. 

This set of steps is repeated for every pressure gridpoint included in our calculation to establish abundances of cloud-forming materials across a wide range of T-P parameter space.  Since we have implicitly assumed that decreasing temperature corresponds to increasing altitude, our method of determining the rainout composition only remains fully generalized for planetary atmospheres with negative vertical temperature gradients (i.e. those without temperature inversions).  For planets with temperature inversions, a more accurate approach is to calculate rainout abundances following the T-P profile vertically upward through the atmosphere rather than using a pre-established T-P grid, as described below.

\subsection{Atmosphere Specific Calculations}

For planets in which the T-P profile is already known or has been pre-calculated, our grid-based approach described above is not necessary.  Instead, the location and composition of condensate clouds can be determined directly along the T-P profile, still employing Gibbs free energy minimization calculations with rainout.  We have additionally developed calculations of this type, both to verify our more general approach of calculating rainout-based condensation curves on a large temperature-pressure grid, and to develop a tool for atmosphere-specific cloud chemistry studies.  The atmosphere-specific version of our code differs from the grid-based version described above in that the atmospheric composition is calculated directly along a vertical trajectory through the planetary atmosphere, and the rainout of cloud-forming species occurs as successive steps of this calculation climb from lower to higher altitude.  This is implemented by starting at the highest pressure grid point of the T-P profile and running the Gibbs free energy minimization and rainout calculations as our code steps down in pressure (and upward in height) through the T-P grid points of the atmospheric profile.  In this framework, the formation of a cloud layer is indicated each time a condensate appears (and is subsequently removed by the rainout calculation).

\section{Results \label{results}}

The results from our grid-based rainout calculations are shown in Figures~1-5 for solar composition and degassed atmospheres arising from CM, H, CV, and EL chondritic meteorites. 
The base of a cloud deck will be located at the lowest point in the atmosphere where a condensate precipitates out of the gas phase.  
For this reason, the composition and location of clouds are determined by finding the intersection point between the condensate's saturation vapor pressure curve and the T-P profile of the planet's atmosphere, provided that this point is characterized by a phase change from gas to solid along an upward trajectory.
In cases where the condensation curve of a particular molecule crosses the T-P profile of the planet more than once (see e.g. Figure~3), the cloud will be expected to form only at the lowest altitude intersection because of the rainout phenomenon.  
Depletion of the cloud-forming materials would not allow additional layers of the same type of cloud to form higher in the atmosphere unless vertical mixing processes such as convection are able to loft the material up to the location where a secondary cloud would form. Atmosphere-specific rainout calculations along the T-P profiles for  GJ~1214b and HD~97658b \citep[][and Jonathan Fortney private communication]{mil10} are included in Figure 6 to further illustrate this phenomenon. 
We note excellent agreement between the location and composition of clouds in our grid-based (Figures~1-5) and atmosphere-specific cloud calculations for both of these planets.

\subsection{Atmosphere Classification and Benchmarks}

We include a solar composition calculation in Figure~1 as a benchmark against other studies that have looked at condensation in giant planet and brown dwarf atmospheres.  
We obtain nearly identical condensation curves (with several notable exceptions) to \citet{Lod06}, who performed rainout calculations for a solar composition mixture.  Mainly, \citet{Lod06} obtained condensates of both forsterite (Mg$_2$SiO$_4$) and enstatite (MgSiO$_3$) at slightly lower temperatures. In our own rainout calculation, enstatite cannot appear once forsterite has already condensed out because the forsterite cloud fully depletes atomic Mg from the atmosphere.  Instead, our calculations predict a cloud layer of SiO$_2$ in place of MgSiO$_3$. The Mg condensation sequence is also discussed in \citet{Vis10}, with the authors concluding that enstatite clouds can form if sufficient vertical mixing takes place.  Otherwise an SiO$_2$ cloud will form deeper in the atmosphere, as in our calculations.  We also find good agreement between our no-rainout calculations (not shown) and those of \citet{1990} and \citet{1999}. 

We find that the degassed chondritic atmospheres can be grouped into four categories and that atmospheres in the same category have very similar condensation behavior.  For this reason, we have only plotted condensation curves for four of the eight meteoritic compositions that we investigated.  The four categories are  (1) reducing atmospheres with sub-solar C:O ratio (R-sub), (2) oxidizing atmospheres with sub-solar C:O ratio (O-sub), (3) reducing atmospheres with super-solar C:O ratio (R-sup), and (4)  oxidizing atmospheres with super-solar C:O ratio (O-sup).  Of the chondritic atmospheres that we study, CM and CI are R-sub, CV is O-sub, H, EH, LL and L are R-sup, and EL is O-sup.  We have listed the C:O and H:O ratios for each of the chondritic atmospheres in Table~2 based on the atomic compositions listed in Table~1. For atmospheres within the same category, the same set of condensates tends to form although the exact location of the condensation curves can be shifted slightly in temperature between models due to small differences in mixing ratios of the cloud-forming materials.  The key cloud condensates that we obtain are reported in Table 3, organized based on the type of atmosphere in which they appear.

\begin{figure}
\includegraphics[angle=90,scale=.4,clip=true,trim= 60 50 50 140]{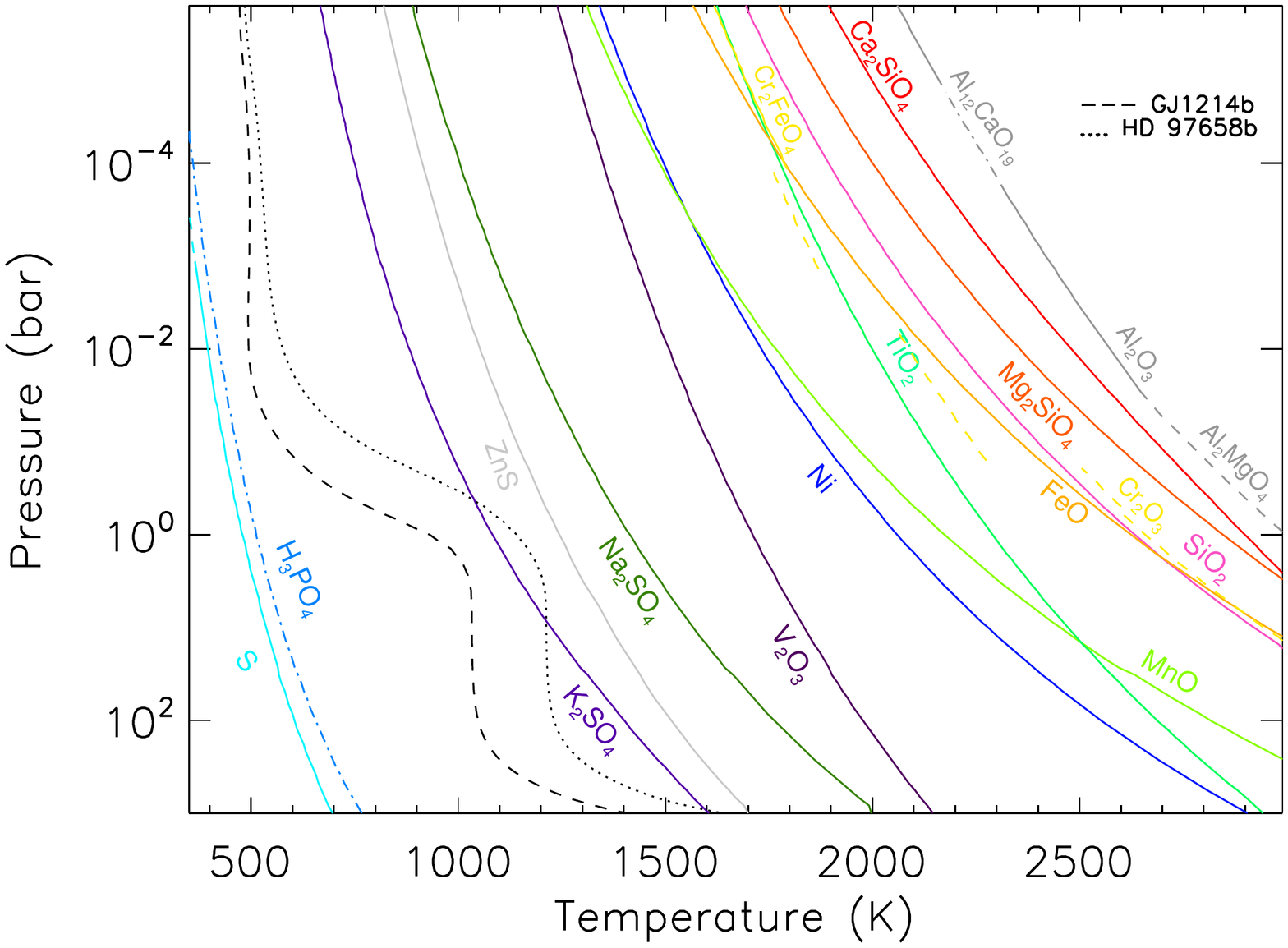}
\caption{Same as Figure~\ref{fig:fig2} but for CV chondritic material.}
\label{fig:fig4}
\end{figure}

\subsection{Condensate Clouds in the Super-Earths GJ1214b and HD 97658b -- Low-Temperature Condensates}
 
At the time of publication of this paper, GJ~1214b and HD 97658b are the only two super-Earths with transmission spectrum observations. 
By comparing the T-P profiles of these two planets to the condensation curves reported in Figures~1-5, we can comment on the composition of potential cloud layers, assuming that clouds are formed by equilibrium processes.  Alternatively, \citet{gj1214b} and \citet{kemp2012} have separately investigated the possibility that the clouds in GJ~1214b's atmosphere could be caused by photochemically induced hazes.

For the solar composition atmosphere, we find that KCl and ZnS are predicted to condense in the atmospheres of GJ 1214b at a pressure of  $\sim$0.5 bar and HD 97658b at a pressure of  $\sim$0.1 bar, in close agreement with previous work \citep{gj1214b, kemp2012}.  The cloud base for these molecules is too deep on its own to explain the featureless nature of GJ~1214b's transmission spectrum, which requires clouds present at a pressure less than 0.01 mbar for solar composition \citep{kreidberg}.  However, \citet{gj1214b} showed that KCl and ZnS clouds could produce a flat transmission spectrum if vigorous vertical mixing with a low sedimentation efficiency (implying a small cloud particle size) is able to carry cloud particles to much greater altitudes.  A second possibility is that inaccuracies in the theoretical T-P profile reported in Figure~1 (caused by e.g.~non-solar chemical abundances, incomplete opacity data, or 3-D effects) could shift the cloud higher in the atmosphere.  Figures~1 and 6 also reveal a secondary deeper cloud layer of Na$_2$S that is expected to be present in the atmosphere of HD~97658b and potentially GJ~1214b.  

As for the composition of clouds in the chondritic atmospheres, we find that it heavily depends on the H:O and C:O ratios. We obtain KCl clouds for both GJ~1214b and HD~97658b for the reducing atmospheres (CM and H meteorites, Figures 2, 3, and 6). 
For the oxidizing atmospheres (CV and EL), we find that potassium is incorporated into condensed K$_2$SO$_4$ instead of KCl and that no clouds appear as shown in Figure 4, 5, and 6. Instead, owing to its higher condensation temperature, K$_2$SO$_4$ is already in the condensed phase when crossing the T-P profile of HD~97658b. There is, therefore, no phase change associated with these intersection points and subsequently no cloud base.
We note however that the T-P profiles plotted in Figures~1-6 were calculated assuming a bulk solar composition atmosphere.  Under the metal-rich conditions of a chondritic atmosphere, the atmospheric temperature is likely higher.  This would in turn shift the location of any cloud layers to higher in the atmosphere and could be responsible for the formation of a K$_2$SO$_4$ cloud layer. The cutoff between potassium forming into KCl vs.~K$_2$SO$_4$ appears to occur near H:O of unity, further supporting the idea that the change in chemistry is tied to the oxidation state of the atmosphere. 

Unlike the solar composition atmosphere, we do not predict any clouds formed of sodium compounds, and we instead find that sodium condenses into Na$_2$S, Na$_2$SO$_4$, and Na$_2$CO$_3$ at somewhat higher temperatures in the metal-rich chondritic atmospheres. As for zinc, we still predict ZnS condensation for outgassed atmospheres of chondritic origin, although this occurs at higher temperatures owing to the larger relative abundance of Zn in these atmospheres relative to solar composition.  In one class of atmosphere (Osup) we also find ZnO condensation at depth.  This results from the high oxidation state of these atmospheres coupled with the removal of carbon from graphite condensation, resulting in excess free oxygen to favor the formation of ZnO.

For the carbon-rich chondritic atmospheres (H and EL), we find that graphite clouds appear at or near the temperature range of GJ~1214b and HD 97658b. The appearance of graphite clouds in atmospheres with high C:O ratios has been discussed previously by \citet{Tar87}, \citet{Sha88}, \citet{Sha95}, \citet{Lod97}, \citet{Mos13b} and \citet{Mos13} for atmospheres of near-solar composition. Specifically, \citet{Sha95} examined the effect of a variable C:O ratio on condensation chemistry.  Their work found graphite typically appears for C:O$>1$. \citet{Mos13b} also explored the influence of both the C:O ratio and atmospheric metallicity on the chemistry of hot Neptunes.  The authors found that, for high metallicity atmospheres, graphite is stable at lower C:O ratios -- down to at least C:O $\approx$ 0.7.  For our own models, based on Figures~2-5, we estimate a cutoff C:O ratio of 0.5-0.6 for the appearance of graphite clouds in chondritic atmospheres.  We attribute our lower cutoff value to the fact that our models include rainout whereas some of the previous studies did not.  Our models also extend to even higher effective metallicity than those of \citet{Mos13b}, which allows for graphite to remain stable at lower C:O values.

\begin{figure}
  \centering
  \begin{subfigure}
    \centering
    \vspace*{-1cm}
    \includegraphics[angle=90,scale=.4, trim= 60 20 10 140]{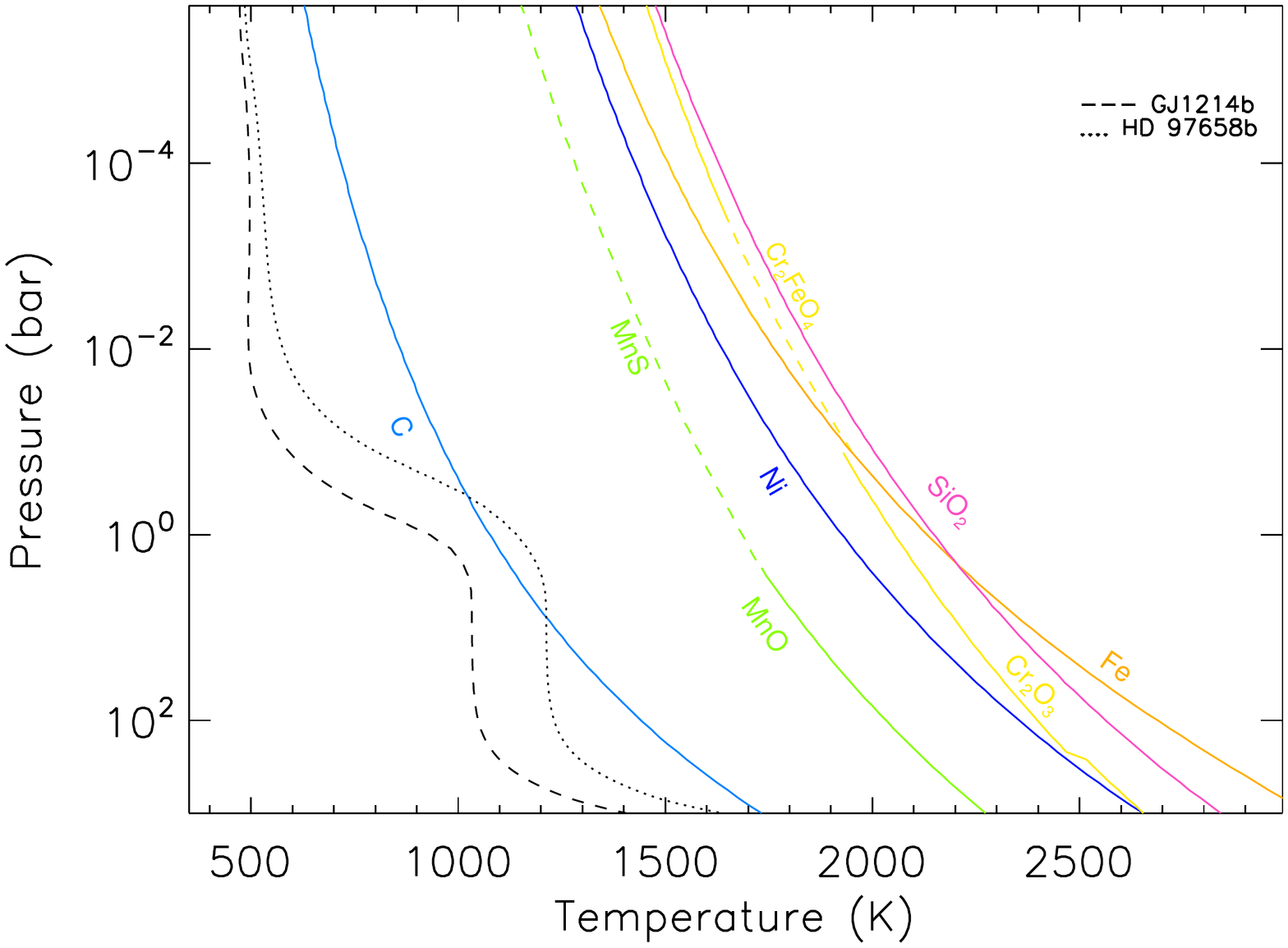}
  \end{subfigure}%
  \quad
  \begin{subfigure}
    \centering
    \includegraphics[angle=90,scale=.4, trim= 20 50 50 140]{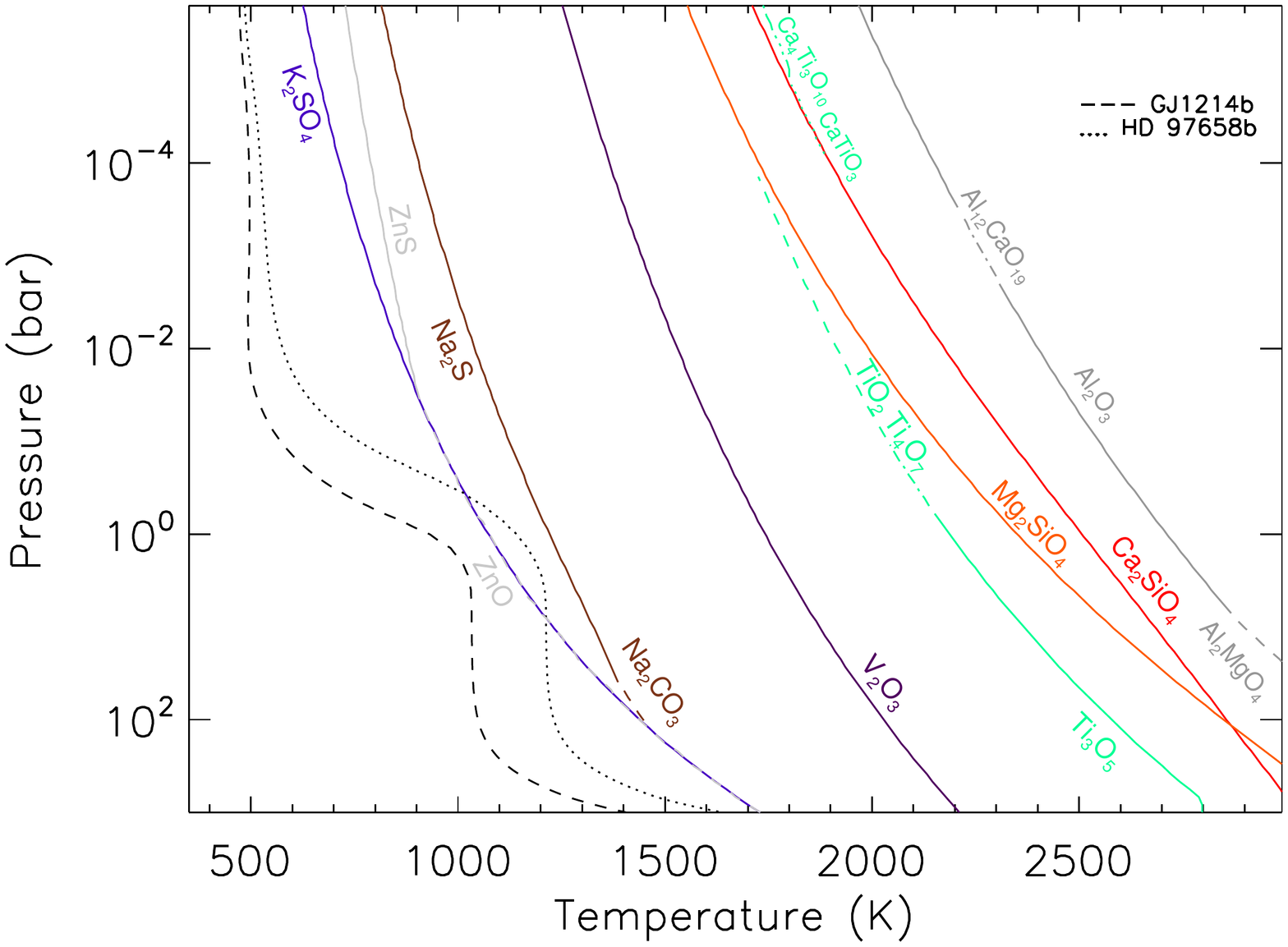}
    \vspace*{-1cm}
  \end{subfigure}
  \caption{Same as Figure~\ref{fig:fig2} but for EL chondritic material.}
  \label{fig:figures2}
\end{figure}

\subsection{High-Temperature Condensates}

Starting from the high-temperature end of our calculations, we find unsurprisingly that refractory elements tend to condense and rain out first.  Condensates of Al, Ca, Mg, Si, and Ti appear first, followed by Fe, Ni, Cr, V, and Mn.  All of these condensates appear at temperatures well above 1,000 K, so they would typically not be present in all but the most irradiated super-Earth atmospheres or potentially in collisionally heated atmospheres for planets that have experienced a recent giant impact \citep[i.e.][]{Lup14, Mil09}.  Below 1,000 K, sodium condensates appear followed by zinc, potassium, and carbon (for C-rich atmospheres) as discussed above.  Ices begin to appear at several hundred Kelvin. 

The vast majority of the high-temperature condensates that we obtain in Figures~1-5 and listed in Table~3 have known relevance both astrophysically and geophysically. Many of these species have been identified in previous equilibrium chemistry cloud studies of exoplanet and brown dwarf atmospheres, including Al$_{2}$O$_{3}$, Al$_{2}$MgO$_{4}$, Fe, MnS, Mg$_2$SiO$_4$ (forsterite), CaAl$_{12}$O$_{19}$ (hibonite), and Ni \citep{1999, Lod06}.  Additionally, TiO$_{2}$, SiO$_{2}$, FeO, and CaTiO$_{3}$ have all been named previously as candidates for dusty cloud grains in giant gas planets \citep{Hel09}.  Ca$_3$Ti$_2$O$_7$, Ca$_4$Ti$_3$O$_{10}$, and CaTiO$_3$ have also been identified as Ti-bearing condensates in dwarf atmospheres \citep{Lod02prime}.
The presence of chromite  (Cr$_{2}$FeO$_{4}$) has been reported in LL chondrites \citep{Kim06}. Dicalcium silicate (Ca$_{2}$SiO$_{4}$) exists on Earth in several polymorphic forms (a few forms are stable at room temperature) and is widely used in the cement industry \citep{Gos79}. Finally, the remaining condensates (e.g.~V$_{2}$O$_{3}$, MnO, C$_{2}$O$_{3}$) are oxidized variants on the condensates that appear under the chemically reducing conditions of a solar composition atmosphere.    

\begin{figure*}
\includegraphics[angle=90,scale=0.78,clip=true, trim= 50 40 0 140]{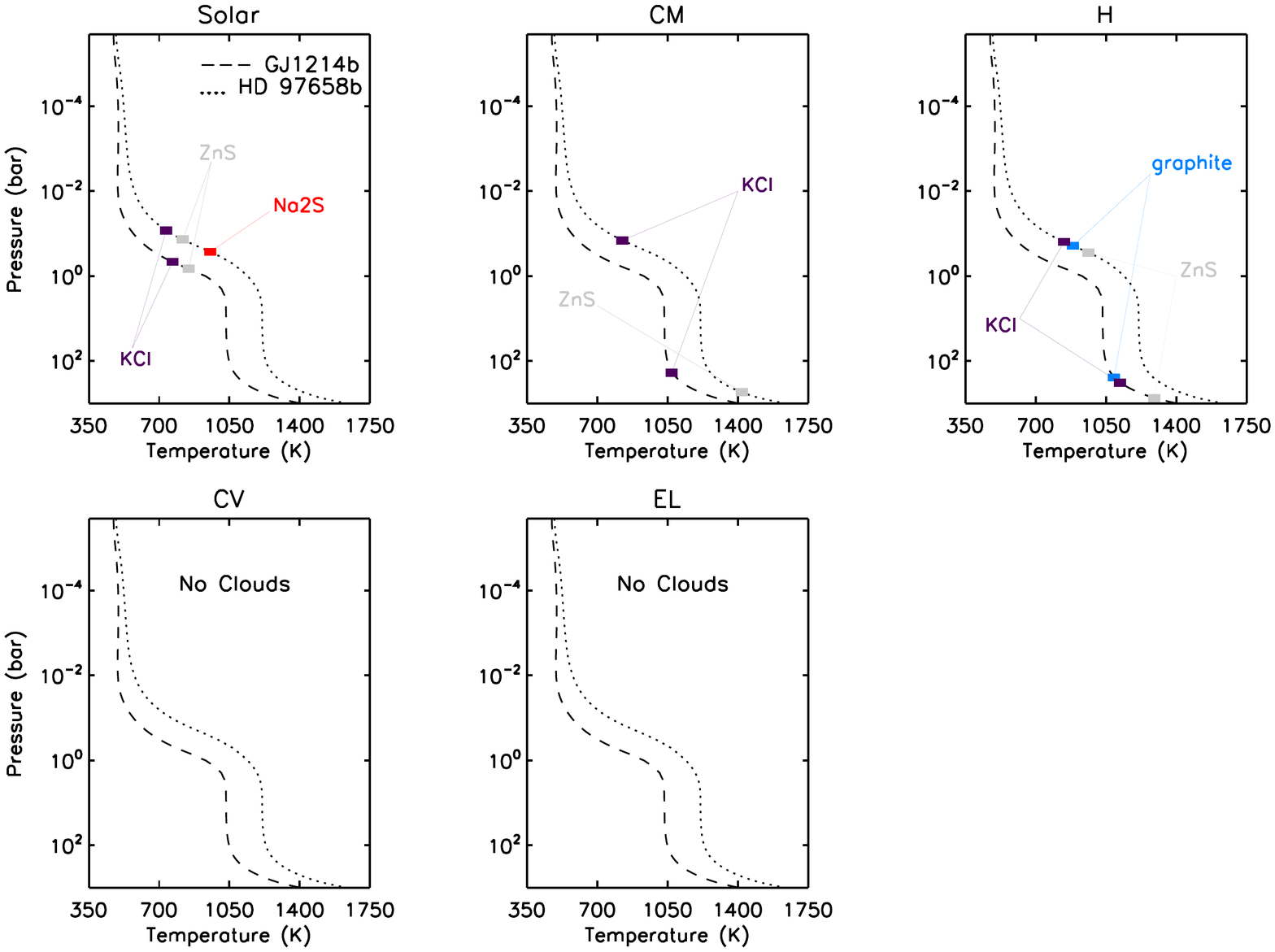}
\caption{Atmosphere-specific rainout calculations along the 1-D T-P profiles for solar composition atmospheres of GJ~~1214b and HD~97658b.  The location and composition of cloud layers are indicated as filled squares.}
\label{fig:fig6}
\end{figure*}

We find that the presence of a number of the high-temperature condensates reported in Table~3 depends strongly on the oxidation state and C:O ratio of the atmosphere in question.  
For example, Fe and MnS appear at super-solar C:O ratios, whereas the oxidized variants FeO and MnO appear in the sub-solar models. Titanium follows a similar pattern where TiO$_2$ is present in the oxidizing and sub-solar C:O models, but this molecule is accompanied or replaced by other less highly oxidized titanium condensates in the super-solar C:O models.  For solar composition (and to a lesser extent the R-sup models) titanium tends to bond with calcium to form even higher temperature condensates.
The sodium condensate chemistry also depends on the oxidation state of the atmosphere.  Sodium condenses as a sulfide, Na$_2$S, in the solar composition atmosphere --- our most reducing case. As the atmosphere becomes more oxidizing, we continue to obtain condensation of sodium sulfide for the H, CM, and CV  chondritic material at lower pressures along with sodium carbonate (Na$_2$CO$_3$), an oxygen compound, at higher pressures.  For the EL chondritic atmosphere, our most oxidizing case, sodium condenses as sodium sulfate (Na$_2$SO$_4$).

The appearance of more highly oxidized condensates at high temperatures in models with lower C:O ratios regardless of the oxidation state of the atmosphere can be attributed to the gas phase carbon chemistry.  Carbon and oxygen bond to form large quantities of CO or CO$_2$.
For atmospheres with low C:O ratios, carbon is the limiting reagent, and oxygen is left over to form other molecules.  However, at high C:O, oxygen becomes the limiting reagent, and  the atom is almost fully consumed in gas-phase CO and CO$_2$, preventing oxygen-bearing condensates such as FeO, MnO, and TiO$_2$ from forming.  

\begin{deluxetable*}{crrrrrrrrr}
\tabletypesize{\footnotesize} 
\tablecolumns{10} 
\tablewidth{0pt} 
\tablehead{  
\colhead{Atom}	&	\colhead{Solar}	&	\colhead{CI}	&	\colhead{CM}	&	\colhead{CV}	&	\colhead{EH}	&	\colhead{EL}	&	\colhead{H}	&	\colhead{L}	&	\colhead{LL} }	\\
H	&	9.11E-01	&	5.23E-01	&	5.32E-01	&	1.24E-01	&	5.60E-01	&	1.89E-01	&	6.19E-01	&	5.55E-01	&	5.95E-01	\\	
He	&	8.78E-02	&	0.00E+00	&	0.00E+00	&	0.00E+00	&	0.00E+00	&	0.00E+00	&	0.00E+00	&	0.00E+00	&	0.00E+00	\\	
C	&	2.65E-04	&	8.04E-02	&	7.11E-02	&	2.47E-01	&	1.74E-01	&	3.68E-01	&	1.49E-01	&	1.81E-01	&	1.49E-01	\\	
N	&	7.31E-05	&	8.20E-03	&	5.70E-03	&	1.00E-04	&	1.32E-02	&	1.85E-02	&	3.73E-03	&	3.33E-03	&	2.90E-03	\\	
O	&	5.30E-04	&	3.77E-01	&	3.80E-01	&	5.91E-01	&	2.44E-01	&	4.20E-01	&	2.23E-01	&	2.55E-01	&	2.46E-01	\\	
F	&	3.15E-08	&	7.18E-07	&	5.85E-07	&	2.88E-06	&	1.78E-06	&	8.31E-07	&	9.33E-07	&	1.04E-06	&	1.21E-06	\\	
Ne	&	8.05E-05	&	0.00E+00	&	0.00E+00	&	0.00E+00	&	0.00E+00	&	0.00E+00	&	0.00E+00	&	0.00E+00	&	0.00E+00	\\	
Na	&	2.16E-06	&	4.91E-05	&	4.00E-05	&	1.97E-04	&	1.22E-04	&	5.68E-05	&	6.38E-05	&	7.11E-05	&	8.26E-05	\\	
Mg	&	3.82E-05	&	8.70E-04	&	7.10E-04	&	3.49E-03	&	2.16E-03	&	1.01E-03	&	1.13E-03	&	1.26E-03	&	1.46E-03	\\	
Al	&	3.15E-06	&	7.18E-05	&	5.85E-05	&	2.88E-04	&	1.78E-04	&	8.31E-05	&	9.33E-05	&	1.04E-04	&	1.21E-04	\\	
Si	&	3.75E-05	&	8.53E-04	&	6.96E-04	&	3.42E-03	&	2.12E-03	&	9.88E-04	&	1.11E-03	&	1.24E-03	&	1.44E-03	\\	
P	&	3.14E-07	&	7.14E-06	&	5.83E-06	&	2.87E-05	&	1.77E-05	&	8.27E-06	&	9.29E-06	&	1.03E-05	&	1.20E-05	\\	
S	&	1.67E-05	&	8.50E-03	&	8.90E-03	&	2.66E-02	&	1.77E-03	&	6.00E-04	&	1.97E-03	&	2.03E-03	&	2.47E-03	\\	
Cl	&	1.96E-07	&	4.47E-06	&	3.64E-06	&	1.79E-05	&	1.11E-05	&	5.17E-06	&	5.81E-06	&	6.47E-06	&	7.52E-06	\\	
Ar	&	3.84E-06	&	0.00E+00	&	0.00E+00	&	0.00E+00	&	0.00E+00	&	0.00E+00	&	0.00E+00	&	0.00E+00	&	0.00E+00	\\	
K	&	1.38E-07	&	3.15E-06	&	2.57E-06	&	1.26E-05	&	7.81E-06	&	3.65E-06	&	4.09E-06	&	4.56E-06	&	5.30E-06	\\	
Ca	&	2.36E-06	&	5.36E-05	&	4.37E-05	&	2.15E-04	&	1.33E-04	&	6.21E-05	&	6.97E-05	&	7.77E-05	&	9.03E-05	\\	
Ti	&	9.08E-08	&	2.07E-06	&	1.69E-06	&	8.29E-06	&	5.12E-06	&	2.39E-06	&	2.69E-06	&	2.99E-06	&	3.48E-06	\\	
V	&	1.08E-08	&	2.46E-07	&	2.01E-07	&	9.87E-07	&	6.10E-07	&	2.85E-07	&	3.20E-07	&	3.56E-07	&	4.14E-07	\\	
Cr	&	4.82E-07	&	1.10E-05	&	8.95E-06	&	4.40E-05	&	2.72E-05	&	1.27E-05	&	1.43E-05	&	1.59E-05	&	1.85E-05	\\	
Mn	&	3.44E-07	&	7.82E-06	&	6.38E-06	&	3.14E-05	&	1.94E-05	&	9.06E-06	&	1.02E-05	&	1.13E-05	&	1.32E-05	\\	
Fe	&	3.14E-05	&	7.15E-04	&	5.83E-04	&	2.87E-03	&	1.77E-03	&	8.28E-04	&	9.29E-04	&	1.04E-03	&	1.20E-03	\\	
Ni	&	1.79E-06	&	4.08E-05	&	3.33E-05	&	1.64E-04	&	1.01E-04	&	4.72E-05	&	5.30E-05	&	5.91E-05	&	6.86E-05	\\	
Zn	&	4.60E-08	&	1.05E-06	&	8.53E-07	&	4.20E-06	&	2.59E-06	&	1.21E-06	&	1.36E-06	&	1.52E-06	&	1.76E-06	\\	

\tablecomments{Atomic abundances for degassed atmospheres from chondritic planetary interiors.  Solar composition is also included as a benchmark using the atomic solar system abundances of \citet{Lodders}. Abundances are normalized to sum to unity.  \label{tab_abunds}}
\label{table:Table 1}
\end{deluxetable*}

\begin{deluxetable*}{crrrrr}

\tablehead{ 
\colhead{Atomic Composition} & \colhead{H-abundance} & \colhead{O-abundance} & \colhead{C-abundance} & \colhead{H:O ratio} & \colhead{C:O ratio} }\\
\textbf{Solar}  &  \textbf{9.11E-01}	&	\textbf{5.30E-04}	&	\textbf{2.65E-04}	&	\textbf{1720.45}	&	\textbf{0.50}	\\ 

\textbf{CM}	&	\textbf{5.32E-01}	&	\textbf{3.80E-01}	&	\textbf{7.11E-02}	&	\textbf{1.39}	&	\textbf{0.18}	\\
CI	&	5.23E-01	&	3.77E-01	&	8.04E-02	&	1.38	&	0.21	\\

\textbf{H}	&	\textbf{6.19E-01}	&	\textbf{2.23E-01}	&	\textbf{1.49E-01}	&	\textbf{2.77}	&	\textbf{0.66}	\\
EH	&	1.74E-01	&	2.44E-01	&	1.74E-01	&	2.28	&	0.71	\\
LL	&	5.95E-01	&	2.46E-01	&	1.49E-01	&	2.42	&	0.60		\\
L	&	5.55E-01	&	2.55E-01	&	1.81E-01	&	2.18	&	0.71	\\	

\textbf{CV}	&	\textbf{1.24E-01}	&	\textbf{5.91E-01}	&	\textbf{2.47E-01}	&	\textbf{0.21}	&	\textbf{0.41}	\\	
\textbf{EL}	&	\textbf{1.89E-01}	&	\textbf{4.20E-01}	&	\textbf{3.68E-01}	&	\textbf{0.45}	&	\textbf{0.87}	\\

\tablecomments{H:O and C:O ratios for all 8 chondritic atmospheres and solar composition. The rows printed in boldface are the ones that are depicted in Figures 1-6 and discussed in detail in the text.}
\label{table:Table 2}
\end{deluxetable*}

We expect the H:O and C:O ratios to be a defining aspect of cloud chemistry (and atmospheric chemistry in general) in exoplanet atmospheres.
However, we find that a few condensates persist for all compositions and do not depend on the H:O or C:O ratios. As shown in Figures 1-5, in all of our models we find the presence of SiO$_2$, Ni, Mg$_2$SiO$_4$, Ca$_2$SiO$_4$, and Cr$_2$O$_3$, along with a variety of aluminum oxides and aluminates.  Additionally, V$_2$O$_3$, MnS, and Cr$_2$FeO$_4$ are present in all of the chondritic atmospheres but are replaced by less oxidized condensates for solar composition presumably due to the far greater H:O ratio in this case.

\section{Conclusions and Discussion}

We have determined the composition of cloud layers for degassed super-Earth atmospheres of chondritic origin.  Previous studies of cloud condensation in exoplanet atmospheres have focused on solar composition and near-solar mixtures.  Here we have moved beyond solar composition to study atmospheres of widely varying oxidation state and C:O ratio, and we have found that in many cases cloud composition depends strongly on both of these parameters.  

\begin{deluxetable*}{lllll} 
\tablewidth{0pt}
\tablehead{
\colhead{Atom}	&	\colhead{Rsub (CM, CI))}	&	\colhead{Rsup (H, EH, LL, L)}	&	\colhead{Osub (CV)}	& \colhead{Osup (EL)} }\\
Al	&	CaAl$_{12}$O$_{19}$, Al$_2$MgO$_4$	& CaAl$_{12}$O$_{19}$, Al$_2$O$_3$, Al$_2$MgO$_4$	&	CaAl$_{12}$O$_{19}$, Al$_2$O$_3$, Al$_2$MgO$_4$	&	CaAl$_{12}$O$_{19}$, Al$_2$O$_3$, Al$_2$MgO$_4$ \\
Fe	&	FeO, Cr$_2$FeO$_4$&	Fe, Cr$_2$FeO$_4$	&	FeO, Cr$_2$FeO$_4$	&	Fe, Cr$_2$FeO$_4$	\\
Mn	&	MnO	&	MnS, MnO	&	MnO	&	MnS, MnO	\\
Mg	&	Mg$_2$SiO$_4$, Al$_2$MgO$_4$	&	Mg$_2$SiO$_4$, Mg$_2$TiO$_4$	&	Mg$_2$SiO$_4$, Al$_2$MgO$_4$	&	Mg$_2$SiO$_4$, Al$_2$MgO$_4$	\\
V	&	V$_2$O$_3$	&	V$_2$O$_3$	&	V$_2$O$_3$	&	V$_2$O$_3$	\\
Ti	&	TiO$_2$	& {Ca$_4$Ti$_3$O$_{10}$, CaTiO$_3$, Mg$_2$TiO$_4$,}	&	TiO$_2$	&	{TiO$_2$, Ti$_4$O$_7$, Ti$_3$O$_5$,} \\
 & & TiO$_2$, Ti$_4$O$_7$, Ti$_3$O$_5$ & & CaTiO$_3$, Ca$_4$Ti$_3$O$_{10}$ \\
Ni	&	Ni	&	Ni	&	Ni	&	Ni	\\
K	&	KCl, K$_2$CO$_3$	&	KCl	&	K$_2$SO$_4$	&	K$_2$SO$_4$	\\
Na:	&	Na$_2$S, Na$_2$CO$_3$	&	Na$_2$S, Na$_2$CO$_3$	&	Na$_2$SO$_4$	&	Na$_2$S, Na$_2$CO$_3$	\\
Ca	&	Ca$_2$SiO$_4$, CaAl$_{12}$O$_{19}$	&	Ca$_2$SiO$_4$, CaAl$_{12}$O$_{19}$, 	&	Ca$_2$SiO$_4$, CaAl$_{12}$O$_{19}$	&	Ca$_2$SiO$_4$, Ca$_4$Ti$_3$O$_{10}$, CaTiO$_3$	\\
 & & CaTiO$_3$, Ca$_4$Ti$_3$O$_{10}$ & & \\
Cr:	&	Cr$_2$O$_3$, Cr$_2$FeO$_4$	&	Cr$_2$FeO$_4$, Cr$_2$O$_3$	&	Cr$_2$O$_3$, Cr$_2$FeO$_4$	&	Cr$_2$FeO$_4$, Cr$_2$O$_3$	\\
Si:	&	SiO$_2$, Ca$_2$SiO$_4$, Mg$_2$SiO$_4$	&	SiO$_2$, Ca$_2$SiO$_4$, Mg$_2$SiO$_4$	&	SiO$_2$, Ca$_2$SiO$_4$, Mg$_2$SiO$_4$	&	SiO$_2$, Ca$_2$SiO$_4$, Mg$_2$SiO$_4$\\
C:	&	none	&	C	&	none	&	C	\\ 
Zn:	&	ZnS	&	ZnS	&	ZnS	&	ZnS, ZnO	\\ 

\tablecomments{Key condensates for chondritic atmospheres with rainout.}
\label{table:Table 3}
\end{deluxetable*}

One motivation for this study has been the discovery of clouds in the atmosphere of GJ~1214b, a transiting super-Earth orbiting an M-star.  Previous equilibrium chemistry studies have predicted the existence of KCl and ZnS clouds in this planet's atmosphere.  Our expanded set of atmosphere models has revealed that K$_2$SO$_4$, ZnO, or graphite clouds are also possible if the atmosphere is oxidizing (for the former two molecules) or has a super-solar C:O ratio (for the latter).  Work by \citet{gj1214b} has previously looked at the optical properties of KCl and ZnS clouds to determine whether these could provide a viable explanation for the featureless nature of GJ~1214b's transmission spectrum.  We have now raised the possibility of graphite, potassium sulfate, or zinc oxide clouds, whose effects on the transmission spectrum have not been explored at this time. 

The use of chemical equilibrium calculations to predict the composition of clouds in exoplanet atmospheres has many precedents in the literature \citep[e.g.][]{1999, Lod02, schaef}.  For solar system planets, equilibrium calculations have been used to identify cloud layers within Jupiter's and Saturn's atmospheres \citep{Feg94}, and they correctly predict the average altitude of water clouds in Earth's atmosphere.  However, there are a number of fundamental limitations to chemical equilibrium calculations, which we summarize here for completeness.  A primary concern is that the minimization of Gibbs free energy does not account for cloud microphysics.  For instance, we have not taken into account the cloud nucleation process because we do not have a good understanding of the physical processes by which rainout occurs.  In cases where the abundance of the condensate material is too low, the cloud may not even form. 
Also, we cannot infer the vertical extent of the clouds from our calculations.  Typically, these poorly understood processes are encompassed in a single sedimentation parameter in 1-D models \citep[see e.g.][]{Ack01}.  Additionally, the atmospheric composition obtained from a Gibbs free energy minimization calculation relies on the list of atoms and molecules that have been supplied as input.  Our database of over 550 molecular species encompasses those with thermodynamic properties reported in the literature formed from the most cosmically abundant atoms, including many molecules that are known to be atmospherically and geophysically relevant.  Incorporating future updates to databases of Gibbs free energy of formation along with an expanded list of atoms and molecules as input to our code would help to further 

improve the accuracy of our results.
The formation of clouds also has an important relationship with the temperature structure of a planetary atmosphere.  The presence of a cloud layer will alter the radiative transport through the atmosphere in several ways: by altering the composition of the atmosphere, by providing an additional opacity source, and by providing a source of latent heat.  All of these effects should ultimately be included in a detailed radiative transfer model, so as to correctly predict the temperature structure of an exoplanet atmosphere.  Here we have provided the key first step of identifying the condensates that will form in atmospheres of diverse H:O and C:O ratios.  

The study of exoplanet clouds is an interesting pursuit in its own right because clouds are indicators of the chemical processes taking place in a planet's atmosphere.  They are also a hindrance -- a high cloud deck can obscure deeper regions of a planetary atmosphere, making it difficult or impossible to determine the overall atmospheric composition.  For this reason, identifying cloud-free super-Earths is of the utmost importance for atmospheric follow-up with JWST, and for those planets found to have cloudy atmospheres additional observational work should be done to uniquely identify the cloud composition.  By identifying super-Earth atmospheres whose T-P profiles are not predicted to intersect with any of the condensation curves presented in this paper for a particular underlying atmospheric composition, we can provide a testable prediction for which planets should remain cloud-free.  

\acknowledgments
We would like to thank Adam Burrows for sharing the Gibbs free energy fits from the \citet{1999} paper with us.  We thank Channon Visscher as our helpful and expeditious referee.  RM acknowledges summer funding from the Grinnell College Mentored Advanced Project (MAP) program.  EMRK and RM acknowledge support from the NASA Planetary Atmospheres program (NNX14AP90A).

\bibliography{ms}

\end{document}